\documentclass[sigconf]{acmart}

\AtBeginDocument{%
  \providecommand\BibTeX{{%
    \normalfont B\kern-0.5em{\scshape i\kern-0.25em b}\kern-0.8em\TeX}}}

\setcopyright{acmcopyright}
\copyrightyear{2018}
\acmYear{2018}
\acmDOI{10.1145/1122445.1122456}

\acmConference[Woodstock '18]{Woodstock '18: ACM Symposium on Neural
  Gaze Detection}{June 03--05, 2018}{Woodstock, NY}
\acmBooktitle{Woodstock '18: ACM Symposium on Neural Gaze Detection,
  June 03--05, 2018, Woodstock, NY}
\acmPrice{15.00}
\acmISBN{978-1-4503-9999-9/18/06}

\graphicspath{{figures/}{pictures/}{images/}{./}}
\newcommand{\etal}{\textit{et al}.}

\usepackage{outlines}
\usepackage{caption}
\usepackage{subcaption}

\begin{document}

\title{Mixing realities for sketch retrieval in Virtual Reality}

\author{Daniele Giunchi}
\affiliation{\institution{University College London} United Kingdom}

\author{Stuart James}
\affiliation{Center for Cultural Heritage Technology\\
\institution{Istituto Italiano di Tecnologia} Italy}

\author{Donald Degraen}
\affiliation{German Research Center for Artificial Intelligence (DFKI) \institution{Saarland Informatics Campus} Germany}

\author{Anthony Steed}
\affiliation{\institution{University College London} United Kingdom}


\begin{abstract}
Drawing tools for Virtual Reality (VR) enable users to model 3D designs from within the virtual environment itself.
These tools employ sketching and sculpting techniques known from desktop-based interfaces and apply them to hand-based controller interaction.
While these techniques allow for mid-air sketching of basic shapes, it remains difficult for users to create detailed and comprehensive 3D models.
In our work, we focus on supporting the user in designing the virtual environment around them by enhancing sketch-based interfaces with a supporting system for interactive model retrieval.
Through sketching, an immersed user can query a database containing detailed 3D models and replace them into the virtual environment.
To understand supportive sketching within a virtual environment, we compare different methods of sketch interaction, i.e., 3D mid-air sketching, 2D sketching on a virtual tablet, 2D sketching on a fixed virtual whiteboard, and 2D sketching on a real tablet.
Our results show that 3D mid-air sketching is considered to be a more intuitive method to search a collection of models while the addition of physical devices creates confusion due to the complications of their inclusion within a virtual environment.
While we pose our work as a retrieval problem for 3D models of chairs, our results can be extrapolated to other sketching tasks for virtual environments.
\end{abstract}




\maketitle

\section{Introduction}
Sketching in Virtual Reality~(VR) allows users to design and create objects in 3D virtual space.
To aid users in the creation of complex 3D designs, existing methods are often supported by a retrieval algorithm capable of finding complex designs based on a simple sketch made by the user by searching a model database.
Common approaches can be divided into methods focusing on gestural interaction~\cite{Deering:1995:HVR, Wesche:2001:FFS} or techniques allowing to freely draw sketches in either 2D~\cite{FonsecaSBMU04} or 3D space~\cite{Giunchi:2018:SIM}.
Gestural interaction techniques are widely used to execute an action as a trigger mechanism or depict a simple trajectory in the design space.
While gestures are generally easy to use, they are usually not suitable for characterizing detailed features of an object.
A dictionary of gestures additionally limit the ability of the user to freely express their desires.
This is especially so for the task of retrieval where flexibility is key to finding the relevant content.
However, both 2D and 3D sketches allow the user to convey complex structures including their details.
These techniques extend the scope of potential designs to a large number of objects within a collection with significant variations in terms of both shape, color and texture.
\begin{figure}[t]
  \includegraphics[width=\columnwidth]{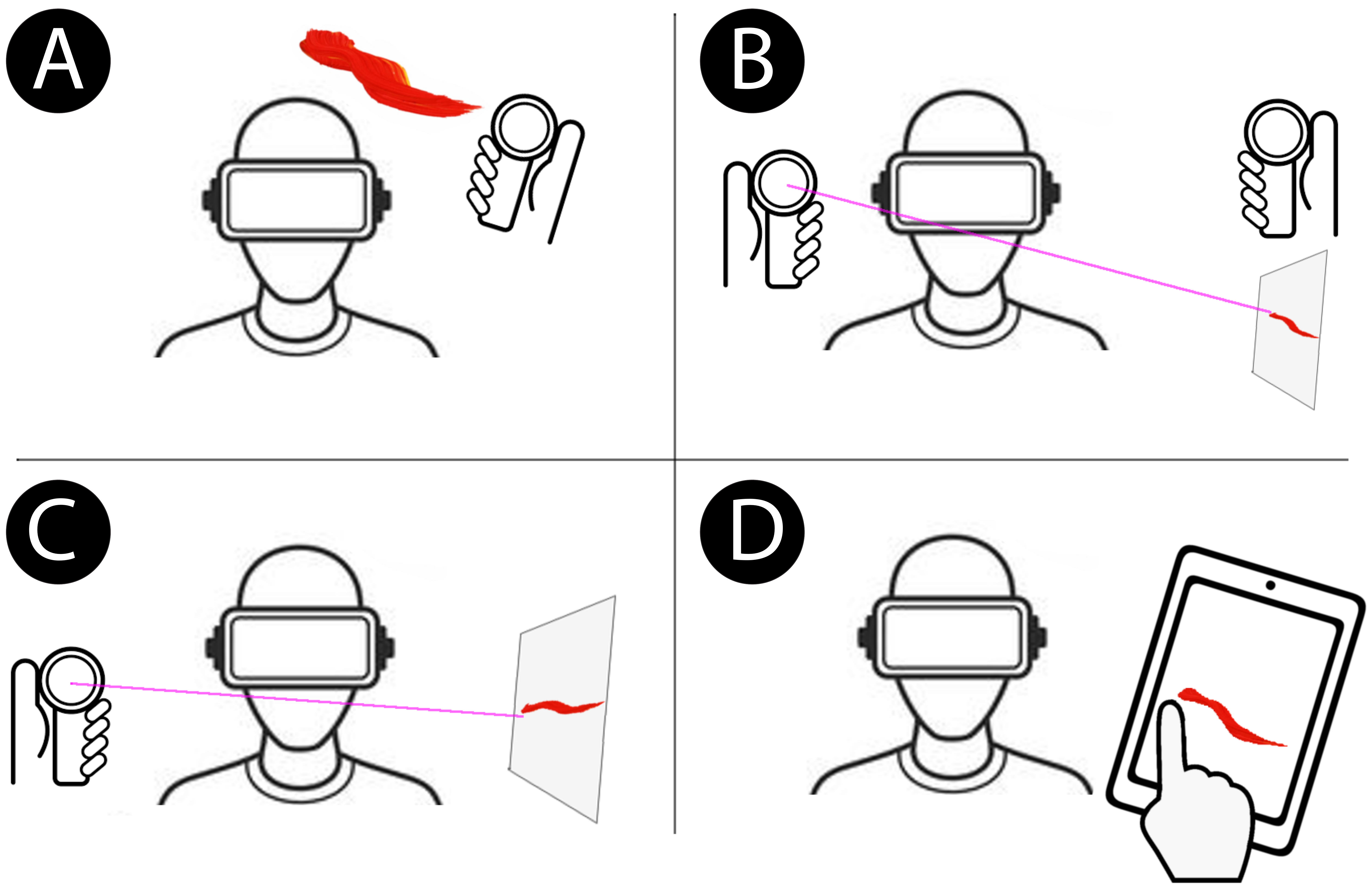}
  \caption{
  To understand supportive sketching within a virtual environment, we investigate sketching in virtual environments and consider 4 different interaction methods,\ i.e.,  (a) 3D mid-air sketching, (b) 2D sketching on a virtual tablet, (c) 2D sketching on a fixed virtual whiteboard, and (d) 2D sketching on a real tablet.}
  \label{fig:methods}
\end{figure}

While 2D drawing is familiar, the extension to 2D sketching of 3D objects is non-trivial where perspective and in turn the users depictive ability plays an important factor. However, understanding whether mouse or pen interaction, \textit{the familiar}, translate into an immersive environment is important to evaluate the contribution of 3D approaches for retrieval and VR interaction in general.

Despite the growing interest in methods for Sketch-based Retrieval~\cite{Hu2010,buiCVIU17compact, DBLP:journals/corr/BuiRPC16,pangCVPR19generalising,DBLP:conf/icip/QiSZL16}, only few examples apply to VR~\cite{Giunchi:2018:SIM, 3dor.20161087}.
The reason for this is twofold: 
(1) the ability for sketch descriptions of an object to adequately express the model, and (2) ability for users to get confidence with the depiction of structural elements.
Prior work by Giunchi~\etal~\cite{Giunchi:2018:SIM} utilized a Multi-View CNN to solve for the first, and introduced on-model sketching in VR for the second. 
In the same study, Giunchi~\etal also performed a detailed study of descriptors for use in 3D model retrieval.

While Giunchi~\etal~\cite{Giunchi:2018:SIM} showed how a combination of sketch and model improves the user's retrieval query, they only explore a single modality of interaction,\ i.e., 3D mid-air sketching, and its level of effectiveness.
However, it is essential to understand how different interaction methodologies can impact on both user performance and user experience. In addition it is crucial to separate the model component from the sketch component because the first one tends to have supremacy over the second one when combined.
Therefore, we present a study to understand how users interact with physical and virtual devices framed in a retrieval context.\\
Our work investigates different techniques for users to provide initial sketch designs as input for sketch-based retrieval algorithms in virtual environments.
Therefore our contribution are as follows:
\begin{outline}
 \1 Four methods of Sketch-based Retrieval interaction in VR:
   \2 3D Mid-Air Sketching, based on the method of Giunchi~\etal~\cite{Giunchi:2018:SIM} using mid-air drawing using a controller;
   \2 2D Sketching on a VR Tablet, using a 2D tablet within the virtual environment;
   \2 2D Sketching on a VR Whiteboard, using a VR plane to annotate the model;
   \2 2D Sketching on a physical tablet, using a real world tablet tracked in VR to annotate the model with strokes.
 \1 An extensive user study over the four methods identifying the advantages of methods with regards to the user.
\end{outline}
In the remainder of this paper we outline related work (sec.~\ref{sec:rel_work}), and detail theon four interaction methods in sec.~\ref{sec:system}.
We provide the details of our user study (sec.~\ref{sec:userstudy}), evaluate our results (sec.~\ref{sec:results}), and summarize our contributions (sec~\ref{sec:conclusion}). 

\section{Related Work}\label{sec:rel_work}
Using \textit{Sketches} for interaction has a long history,~e.g., in the context of the retrieval of images~\cite{BimboPAMI97elasticsketch,EitzSBIM09, Hu2010} or videos~\cite{James2014,HuMMM12}, interaction in 2D~\cite{FonsecaSBMU04}, or more recently for interaction in 3D~\cite{Giunchi:2018:SIM}. 
Given the multi-modal interaction style there is a vast amount of literature that can be considered relevant. We therefore focus on core techniques in 2D sketching (sec.~\ref{sec:rel_2d}) and 3D sketching in AR/VR (sec.~\ref{sec:rel_3d}).

\subsection{2D Sketching for Retrieval}\label{sec:rel_2d}
Sketch-based Retrieval techniques using 2D sketches query rely on a predefined database to obtain either a resulting image (SBIR) or video (SBVR).
SBIR approaches can be classified into blob-based techniques that try to extract features for shape, color or texture for the blob, or contour-based techniques that characterize the image with curves and lines.
Blob-based SBIR methods try to describe images using descriptions of the segments within the image, for example, QBIC~\cite{AshleyFHLNP95}, which creates separate descriptors for the modalities.
Alternatively, topology models~\cite{DBLP:journals/vlc/SousaF10}  can describe the blob characteristics.
Contour-based methods include elastic matching~\cite{DBLP:conf/vl/BimboPS94} and grid and interest points approaches~\cite{DBLP:journals/tsmc/ChalechaleNM05}.
SBVR approaches mainly address the optimization of objects between video and sketch query. Collomosse \etal ~\cite{DBLP:conf/icpr/CollomosseMW08} applied Linear dynamic system to solve the object matching. Hu \etal~\cite{HuMMM12} introduced a hybrid technique that takes advantage of object semantics. While, James \etal~\cite{James2014} proposed a descriptor-based technique for fast indexable retrieval.  To improve on early methods for sketch-based retrieval, more recent literature started to employ neural networks.
An early example was SketchANet~\cite{DBLP:journals/corr/YangH15}, which performed sketch recognition using Alexnet~\cite{krizhevsky2012imagenet}.
More recently, triplet convolutional neural networks have gained interest as they have the capacity to deal with deep embedding spaces~\cite{buiCVIU17compact}. 
Improving the image similarity metric is a main challenge as these triplet architectures are used to measure similarities between images and sketches~\cite{DBLP:journals/corr/BuiRPC16}.
Examples include the work of Wang \etal~\cite{DBLP:journals/corr/WangSLRWPCW14}, where they developed a learning model based on triplets to describe the images considering fine-grained similarities.
This approach was extensively used in recent works like in Wang \etal \cite{DBLP:journals/corr/WangKL15} where the embedding spaces of sketches and 3d models were joint, and in Li \etal~ \cite{DBLP:journals/tog/LiSQFCG15} for images and 3D models.

All the aforementioned approaches focus on 'Drawing on a whiteboard' with generally singular or interactive search. While improvements in the field of model retrieval increase accuracy, supporting the user during sketching is imperative to increase performance.
One example is ShadowDraw~\cite{DBLP:conf/visual/SciascioMM99}, where the user is provided with feedback related to texture, color, and shapes to improve image retrieval task.
During a sketching activity, the system provides the user with a real-time shadow image to help free-form drawing and achieve a better final sketch of the object.
iCanDraw~\cite{DixonCHI10iCanDraw} helps to solve drawing issues by exploiting algorithms for sketch recognition.
Lastly, Sketch-to-Collage~\cite{DBLP:conf/siggraph/RuizSN07} uses a query-by-sketch to generate collages from an image collection and deploys an indexing mechanism based on colors and shapes.

With the rapid growth of 3D model collections the retrieval of a 3D model from a collection has received attention. However, the nature of the sketch is abstract and iconic, and it does not easily lend itself to describing an object in its details from a 2D query. However for image retrieval two classes of descriptors were introduced: model-based descriptors and view based descriptors.
For model-based there consists three types of descriptors: geometric moment \cite{topology_invariant_similarity}, surface distribution \cite{shape_distribution} and volumetric descriptors \cite{DBLP:conf/3dor/Rustamov10}. The common goal for all of them is to extract the features that describe the shape of the objects.
View-based extracts features from a collection of the 2D projections of the 3D models. Ansary \etal ~\cite{bayesan_3D} 
implement an indexing method that exploits 2D views and Su ~\cite{multiview_CNN} implemented a stack of convolutional neural network that takes multiple 2D projections of the 3D model generating a single compact descriptor for the object.

\subsection{3D Sketching in AR/VR}\label{sec:rel_3d}
Rather than transforming 2D sketches into 3D representations, immersive modeling environments allow users to sketch and design 3D content using mid-air interaction techniques.
Such techniques utilizing 3D space are intuitive to learn regardless of the user's expertise with the VR system~\cite{Wesche:2001:FFS}.
A very early example of such an approach called HoloSketch~\cite{Deering:1995:HVR} combined head-tracked stereoscopic shutter glasses and a CRT monitor with a six-axis mouse or \textit{wand} for mid-air freehand drawing.

While 3D freehand drawing can support expert 3D artists~\cite{Schmidt:2009:EPC} and shows improving accuracy and uniformity of sketched objects over time~\cite{Wiese:2010:ILI}, interaction techniques for sketching in 3D are still limited in terms of accuracy and user fatigue.
Realizing an intended stroke during sketching is firstly bound by the user's perception of depth~\cite{Arora:2017:EES}.
When users are unable to determine their active drawing location and spatial relationships with other contents inside the scene, sketching errors occur.
Additionally, the added complexity of 3D compared to 2D sketching causes a higher cognitive load and increasingly taxes the sensorimotor system~\cite{Wiese:2010:ILI}.
Lastly, as the absence of a physical resting surface leads to fatigue, accuracy is negatively influenced accuracy over time~\cite{Arora:2017:EES}.

3D freehand sketching approaches can be further classified in categories that highlight the motivation as well as the limitation of such type of interaction: metaphors and both virtual and physical support surfaces.

For \textit{Sketch metaphor} Wacker \etal \cite{wacker:2019:arpen} designed and implemented a tool called ARPen, whose real-world position is tracked by a smartphone app, that lets the user interact via mid-air sketch. In addition, they evaluated through a user study different techniques to select and move virtual objects with such a tool. 
3D sketching is used in the work of Giunchi \etal \cite{Giunchi:2018:SIM} where the task is retrieving a target model from a large collection of chairs combining sketch input with 3D models by using a multi-view CNN to improve accuracy during the search. A user study where a searching task is proposed to the participants compared the search based on sketch interaction with simple browsing of the chair database.
We develop our study considering the 3D sketch method based on the same freehand interaction but focusing specifically on the aspect of sketch generation.



In recent years in VR environments, the use of 2D surfaces,\textit{Virtual surfaces}, has been explored to replace the inaccuracy brought by 3D tracking. Despite the lack of one dimension some tasks are suited for a 2D device such as terrain editing~\cite{DBLP:journals/cga/BowmanWHA98}, user interface editing~\cite{DBLP:conf/vr/BowmanW01} or even object selection and manipulation~\cite{DBLP:journals/cgf/SzalavariG97}.
Arora \etal developed SymbiosisSketch AR~\cite{Arora:2018:SCS} that combines 3D drawing in mid-air with freehand drawing on a surface. They equipped the user with an Hololens, a tracked stylus, and a tablet and created a hybrid sketching system suitable for professional designers and amateurs.
Machuca \etal supported 3D immersive sketching using multiple 2D planes to create 3D VR drawings~\cite{Machuca:2018:MAF}.

A virtually rendered tablet in VR can not provide the same latency-free response that a 2D tool or \textit{Physical surfaces} can have, different attempts of integrating a real physical 2D tablet have been made.
Arora compared traditional sketching on a physical surface to sketching in VR, with and without a physical surface to rest the stylus on~\cite{Arora:2017:EES}.
Additionally, they investigated visual guidance techniques to devise a set of design guidelines.
Wacker \etal \cite{Wacker:2018:PGA} studied how accurately visual lines and concave/convex surfaces let users draw 3D shapes attached to physical virtual objects in AR and Dorta \cite{doi:10.1177/1478077116638921} draws on virtual planes using a tracked tablets.

\section{Sketch Modalities for VR}
\label{sec:system}

In the following section, we describe our implemented model retrieval system, our four sketch interaction methods and our back-end which acts as retrieval system. For each method, the user is immersed within a virtual environment and sketches either in 3D mid-air (3D Sketching), on a virtual tablet, a virtual whiteboard or on a tracked physical tablet.

\subsection{Interaction Methodologies}
We propose four distinct methods of interaction, 3 of them use 2D sketch generated on different canvas and with different actions, and only one of them make use of 3D sketch. We outline them in detail below.
 
\begin{figure}[tt]
    \centering
    \begin{subfigure}[t]{0.48\linewidth}
        \includegraphics[width=1\columnwidth]{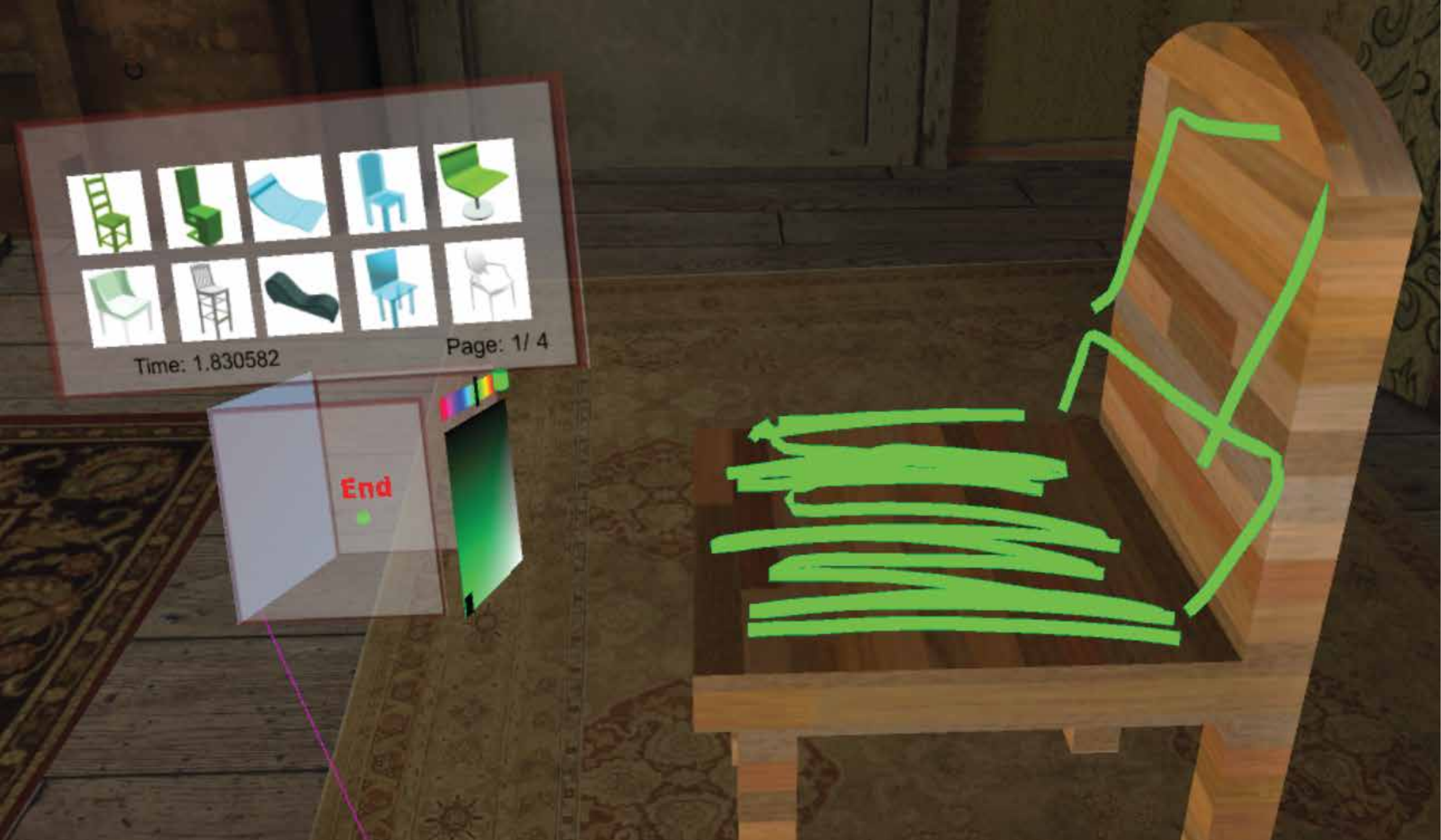}
        \caption{3D Mid-Air Sketching}
        \label{fig:3DSketch}
    \end{subfigure}
    ~
    \begin{subfigure}[t]{0.48\linewidth}
        \includegraphics[width=1\linewidth]{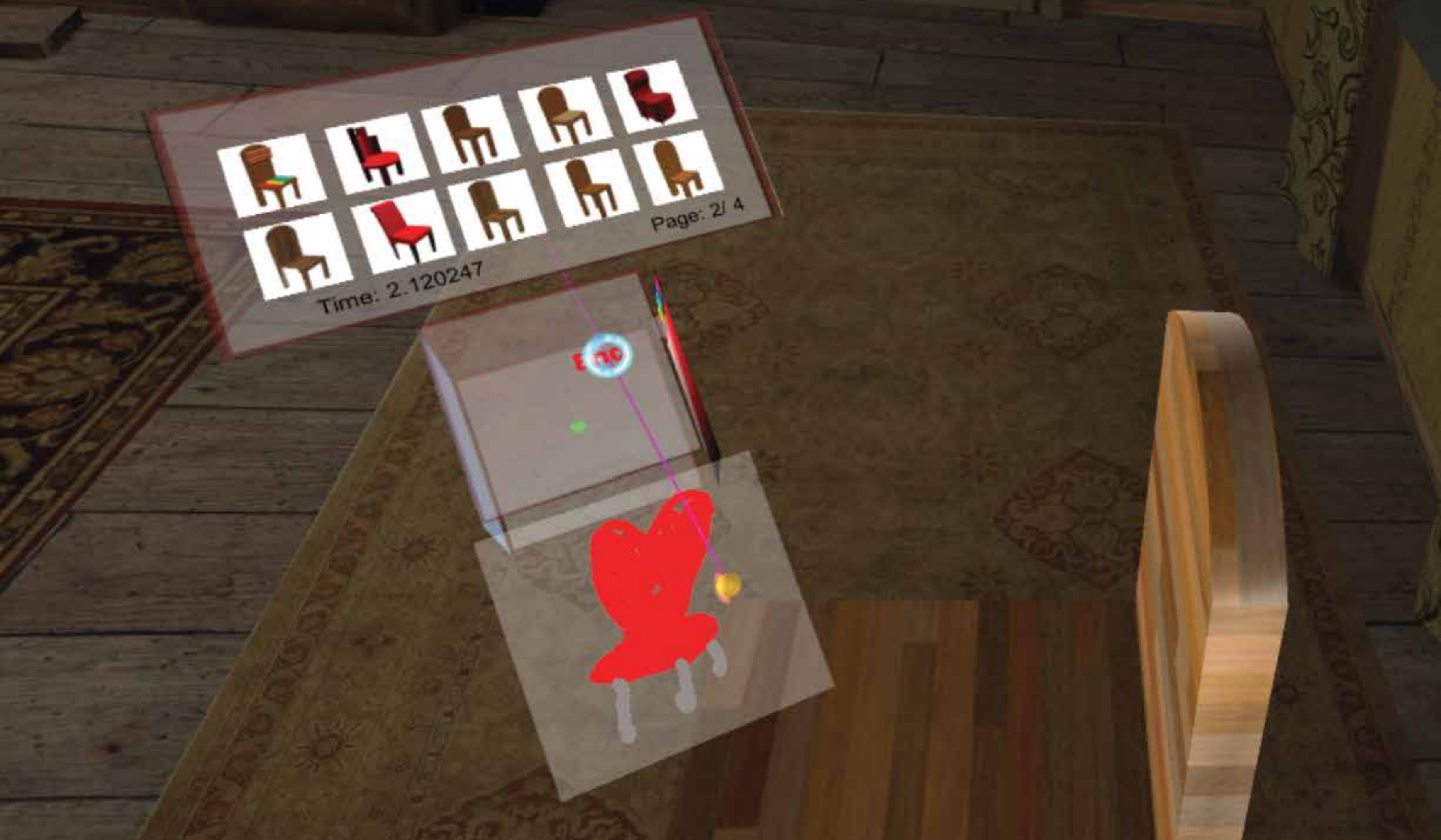}
        \caption{2D Sketching on a VR Tablet}
        \label{fig:vrtablet}
    \end{subfigure}
    \hfill
    \begin{subfigure}[t]{0.48\linewidth}
        \includegraphics[width=1\linewidth]{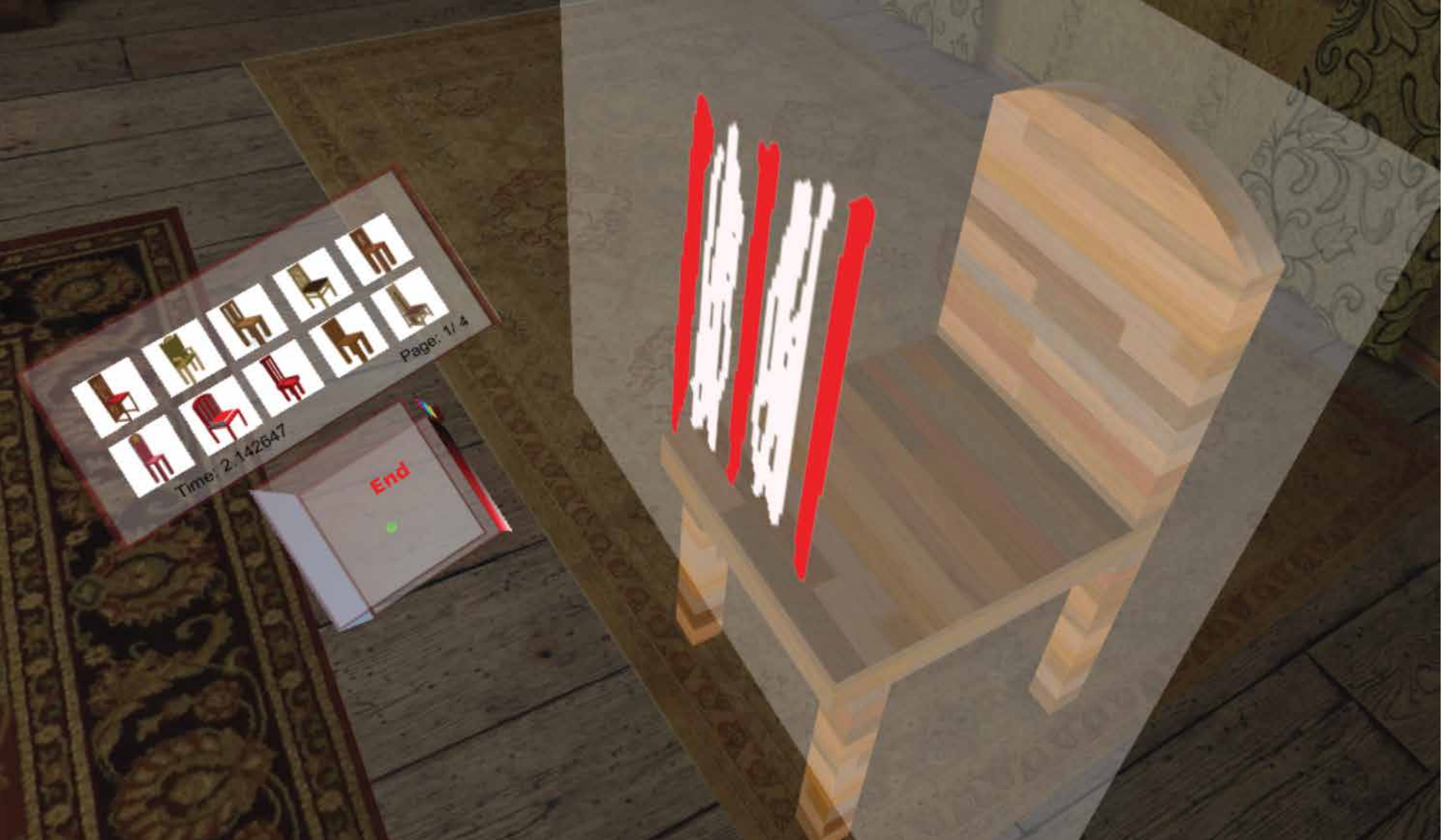}
        \caption{2D Sketching on a VR Whiteboard}
        \label{fig:WhiteBoard}
    \end{subfigure}
    ~
    \begin{subfigure}[t]{0.48\linewidth}
        \includegraphics[width=1\linewidth]{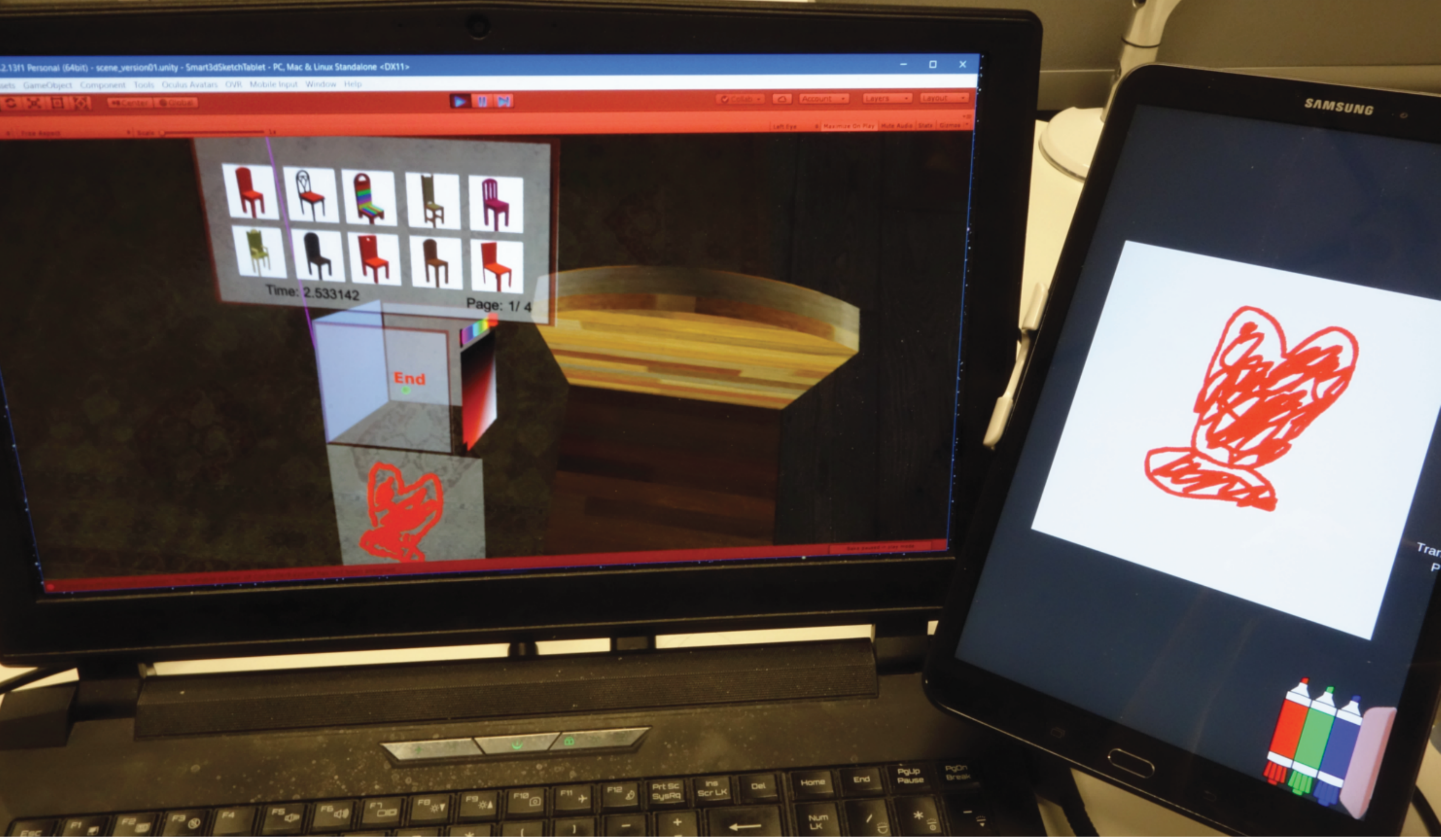}
        \caption{2D Sketching on a Physical Tablet}
        \label{fig:realtablet}
    \end{subfigure}
    \caption{Overview of the four implemented interaction modalities for sketch-based retrieval.}\label{fig:methods2}
\end{figure}

\paragraph*{\textbf{3D Mid-Air Sketching}}
This method, shown in \autoref{fig:3DSketch}, is similar to existing systems for sketching in VR and is directly based on the method for 3D mid-air sketching described in~\cite{Giunchi:2018:SIM}.
The user directly sketches in 3D space using a hand-held controller.
While holding down the trigger button, a virtual stroke is applied in the air at the current position.
By dragging the controller through the air, strokes are extended towards a continuous line.
Once the trigger button is released, the active stroke is considered to be completed, and a new stroke can be initiated.
There is no theoretical limit to the volume the sketch can occupy and the user can create the sketch in any position.


\paragraph*{\textbf{2D Sketching on a VR Tablet}}
With this method, we mimic a natural method of sketching, but placed within VR.

A 2D panel is attached to the user's non-dominant hand controller, see \autoref{fig:vrtablet}, referencing the familiar painting palette. We avoided a user movable panel due to the additional complexity of modelling an unconstrainted position and orientation.
As this method aims to simulate sketching on a portable tablet, we designed the 2D panel with a similar size to a commonly used tablet device (Galaxy Tab A 10.1") with the largest dimension as side of the squared panel.
The actual sketching of lines is done using the controller in the user's dominant hand.
This makes the interaction technique a bi-handed approach as both hands are involved in the process of sketch creation,\ i.e., one hand performs the sketch while the other hand stabilizes the drawing canvas.
Here, the 2D sketch is not only limited in the third dimension, but also by the size of the panel.

\paragraph*{\textbf{2D Sketching on a VR Whiteboard}}
Similar to VR Tablet, the whiteboard method provides a panel onto which the user can sketch in 2D, see \autoref{fig:WhiteBoard}. A familiar design paradigm, the whiteboard technique extends the size of the tablet to that of a larger whiteboard in order to provide more space for sketching. 

The dimension of the whiteboard is linearly five times than the virtual tablet, with the same pixel resolution.
As the whiteboard is positioned on a fixed location (centre of the room) inside the virtual environment, this method does only requires the use of the user's dominant hand. The user could ask, before starting the session to adjust the position of the whiteboard.


\paragraph*{\textbf{2D Sketching on a Physical Tablet}}
Using a real-world tablet (Galaxy Tab A 10.1") approach offers the user a physical tablet to perform 2D sketching while immersed in the virtual environment, see \autoref{fig:realtablet}.
This mimics the most commonly technique used by digital artists.

The tablet is positioned on a table and requires a short registration procedure before the start of a sketching session.
While the tablet is still limited in drawing space, the physical feedback provided from the actual device aims to improve the stability during sketching.
The user is able to sketch using her finger, thus this approach does not require the use of a controller.
The user's hands are tracked using a LEAP motion device as the virtual environment needs to visualize the correct position of the hands of the user.
This additional tracking is necessary as we noticed during a first implementation that the absence of the visual feedback for the finger position led to an unpleasant experience.
This was mainly due to the user being unable to find the right location of contact between her finger and the tablet.

\subsection{Sketch and 3D collection}

The 3D sketch is implemented as a colored strip with a wider section that is exposed to the current camera.
Consequently, all the virtual cameras involved in the multi-view generation render the sketch-oriented along the wider section independently from the length or the trajectory of the stroke. The user can change the color of the next sketch using the palette without modifying the strokes already generated.
The 2D sketches are simply implemented as sequences of connected points on a canvas.

As chairs database we used a set of 3370 chairs that are part of ShapeNet \cite{shape_net}. ShapeNet is a large
collection of 3D models where an extensive subset is made by chairs with or without colors or textures.

\subsection{Model Retrieval}
\label{sec:model_retrieval}
\begin{figure*}[t!]
    \includegraphics[width=\textwidth]{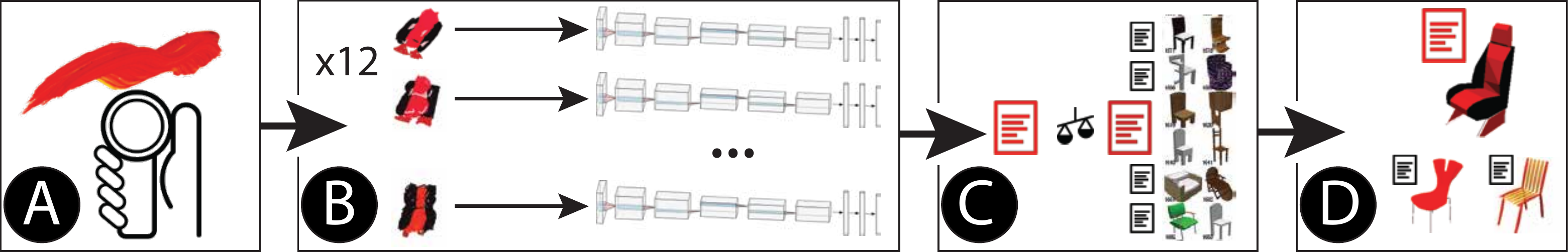}
    \caption{Overview of the system's model retrieval mechanic.
    Here, (A) the sketch created by the user results in a total set of 12 images (B) which are processed by 12 versions of the same CNN.
    After a max-pooling procedure, one descriptor is generated and (C) compared through Euclidean distance with the descriptors previously calculated for all the chairs of the collection.
    The result of the search is (D) a small subset of the most similar chairs from which the user can select.}
    \label{fig:system_overview}
\end{figure*}


To perform sketch retrieval, we implemented a back-end which hosts a pre-loaded CNN model.
This back-end answers the visual queries containing the sketches by producing a list of the 40 models that are considered more similar to the input sketch.
We used the VGG-M Matlab implementation of Su ~\cite{multiview_CNN}.
This implementation provides a single visual descriptor after elaborating the snapshots taken by VR software.
The method works by generating a set of structured camera views (snapshots) around the models. These images are then passed through the CNN and a final descriptor is generated.
During the generation of the snapshots, the cameras in the virtual environment move at different angles looking at the center mass of the sketch and fit in the image its maximum extent.
The CNN uses the VGG-M model~\cite{Chatfield14} which has $5$ convolutional layers and $3$ fully connected ones.
The encodings from each view are then merged through view-pooling using a max function over the views, using the final $3$ fully connected layers of VGG-M to create a descriptor.
We removed the final layer of the model described by Su~\etal, as it is normally used for classification purposes.
In contrast, for the other methods of interaction we bypass the pooling step.
An overview of the entire system is depicted in \autoref{fig:system_overview}.
\section{Experiment}
\label{sec:userstudy}
To compare our four interaction methods for sketching in VR, we designed and conducted a user study performed in our lab. \\\\
\noindent{\bf Participants:}
We used a within-subjects experimental design to help to reduce the number of participants and errors associated with individual differences.
To counterbalance possible carryover effects, the methods were randomized between the users.
As our independent variable, we distinguish the methods used to sketch, 3D sketch, 2D sketch on a virtual tablet, 2D sketch on a fixed virtual whiteboard and 2D sketch on a real tablet.
We distinguish $3$ dependent variables, namely the success rate, the completion time of the task and the number of submitted queries during the search.

A total of 5 participants ($4$ male, $1$ female, $25 - 43$ age range with avg. $34$) volunteered for our study that was approved by ANON ethics board. 
All participants had previous experiences with VR and already used an Oculus RIFT and Touch.
We opted for a within-subjects study with a high number of searching tasks to minimize the possible fluctuations in drawing skills emerging from a study with many users on few chairs. A total of 32 searches were performed by each user, 8 for each of the 4 methods. In total we have 160 searching sessions.
The users did not declare to have particular abilities in drawing or sketching, and they were compensated for the test.\\

\noindent{\bf Apparatus:} 
The rendering of the Virtual Environment was performed in Unity 2018.2.13 using an Oculus RIFT DK1 headset with a connected  laptop computer. 
The specification of the laptop is: Intel i7 CPU, 64 GB RAM with Nvidia GeForce GTX 980M graphics card.
The interaction with the 3D environment was provided by both the Oculus Touch, i.e., two controllers paired with the headset, and hand-tracking using a LEAP Motion device. 
For the real tablet session, we use a Galaxy A6 tablet, tracked within Unity application via an Oculus Touch controller attached on the top right corner with Oculus Rockband VR Touch Guitar Mount Attachment.\\

\noindent{\bf Procedure:}
Before starting the experiment, each participant signed a consent form and was instructed on the searching task.
A period of 15 minutes was dedicated to training the user to develop confidence with the controllers, the sketch mechanism and the virtual environment. Between each method, users had 3 minutes of rest and they can perform the task seated or standing up.

Upon completion of the introduction, the experiment commenced.

Inside the virtual environment, users are placed inside a virtually furnished room with a chair model in the center.
The user searches by sketching and then triggering the search system when he or she wants. Results are displayed on a floating panel. 
The panel is positioned on the left controller, and the selection is made via the right-hand controller.
The panel can display 10 chairs at a time, and 40 chairs are the result of each query.
Therefore the user can navigate 4 pages with simple controls. 
We decided to display only ten models per page in order reduce occlusion of the scene while providing enough variety for the user to choose from.
\\

For each method, participants were asked to perform sketch searches for a given set of $8$ different chairs in shape and textures.
For each session, the participant started with a randomly selected sketch interaction method and performed the search for each target chair of the $8$ proposed in a random order. 
Using the selected method, the participant started sketching to initiate the search for the presented target chair.
Upon confirmation, the system provided the user with a set of potential chairs considered to be most similar to the created sketch.
The participant could refine the search results by editing or detailing the sketch. 
When the participant was satisfied with the results, one of the chairs from the proposed set was selected.
This selection concluded the search task and would replace the presented 3D model inside the scene.
Each session is given a time limit of 3 minutes after which the search was considered terminated without a successful result.
Additional functions available to the participants during sketching with all the different methods were: \textit{undo} function to erase the sketch either entirely or partially, and a \textit{color palette} to characterize the sketch with a chosen color.

The experimenter recorded success rate and completion time for each task.
The user's experience with the current sketching method was measured asking the user to rate the interaction on a scale from 1 (very bad) to 5 (very good).


\section{Evaluation}\label{sec:results}

\begin{figure*}[t]
    \includegraphics[width=\textwidth]{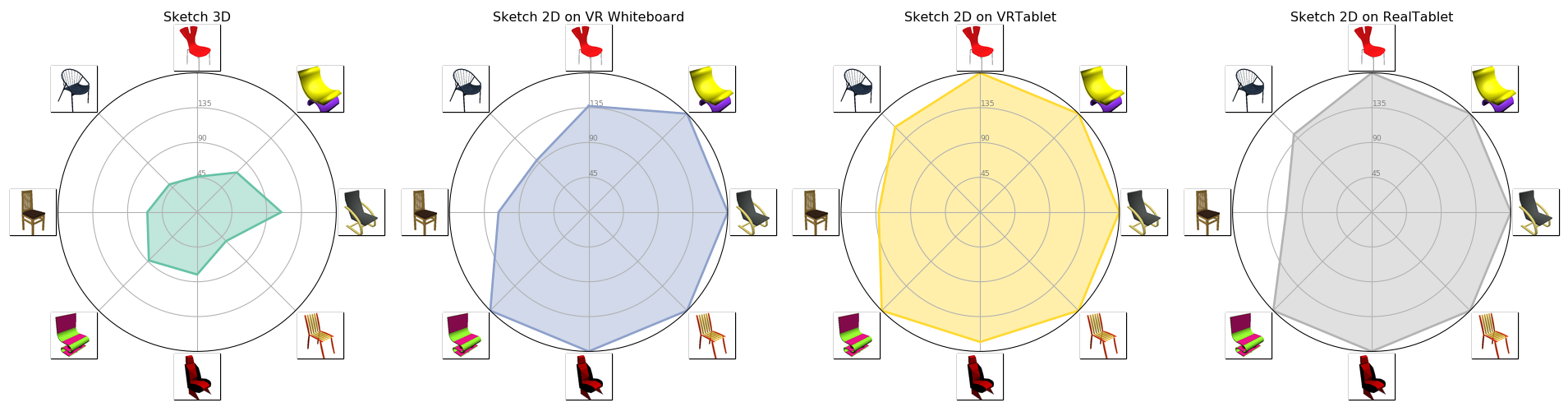}
    \caption{The inner circles in each radar represent 45 seconds. The center of each circle corresponds to time 0. Each radar shows the average time to complete the task for each chair considering all the methods. The time is normalised to $3$ minutes as the upper limit allowed for a search attempt.}
    \label{fig:radar_chart_rs}
\end{figure*}
\begin{figure*}[t]
    \includegraphics[width=\textwidth]{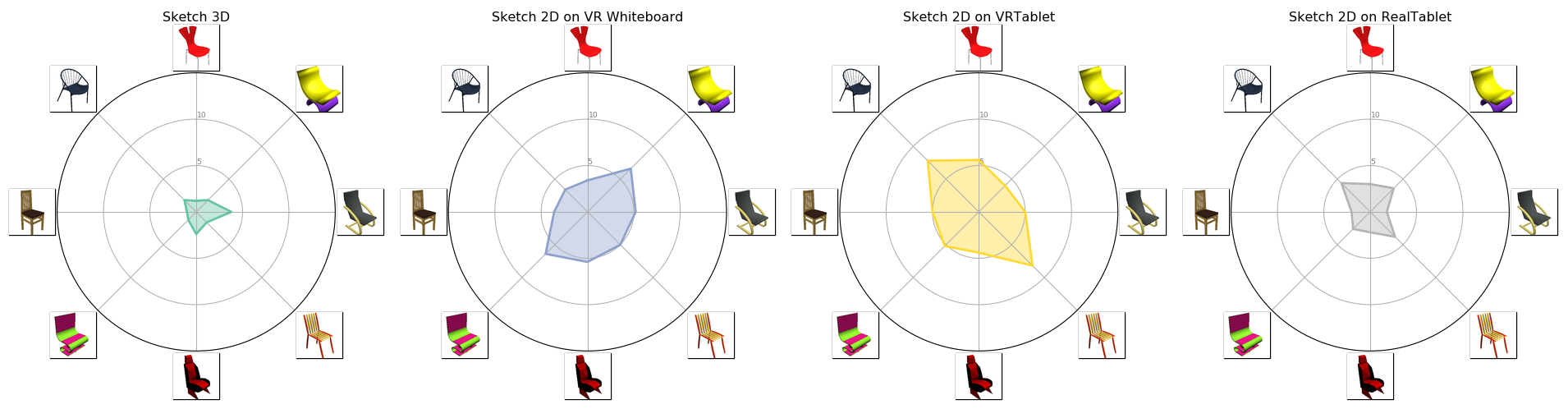}
    \caption{The number of search iterations for the different types of chairs for the different methods of interaction.}
    \label{fig:radar_chart_tries}
\end{figure*}

We investigated the differences between the four different methods of interaction using sketches within a virtual environment.
We evaluated our study over the $8$ distinct chairs presented to each participant and $4$ methods to test in terms of the accuracy of the returned model, time, and the number of queries to complete the task. 
To evaluate the accuracy we counted the number of successful searches among the total number of searches.
The number of successful task completions for the 3D sketch was 37 out of 40 ($92.5\%$), the 2D sketch with the whiteboard was 9 out of 40 ($22.5\%$), the 2D sketch with virtual tablet was 6 out of 40 ($15\%$) and the the 2D sketch with real tablet was 5 out of 40 ($12.5\%$).
To evaluate the efficiency we measured the time elapsed from the beginning to the end of the search and count for each search the number of attempts to submit the sketch to the system.
The average time among all the chairs for the the 3D sketch was 71 seconds, the 2D sketch with the whiteboard was 156 seconds, the 2D sketch with virtual tablet was 169 seconds and the the 2D sketch with real tablet was 166 seconds. 
The average number of attempts among all the chairs for the the 3D sketch was 1.85, the 2D sketch with the whiteboard was 4.88, the 2D sketch with virtual tablet was 5.65 and the the 2D sketch with real tablet was 2.9. 
We demonstrated how the variation in chairs effects the different methods in the radar plot of Figure~\ref{fig:radar_chart_rs}. 

We showed in Figure~\ref{fig:triggerCounts} the cumulative number of attempts over the different methods of interaction. It can be seen that the 2D virtual tablet required a significant numbers of search triggers, while the 3D sketch required the least. In Figure~\ref{fig:radar_chart_tries}, we show how the difference in 3D model effects the number of required search triggers. These results mimic those of fig.~\ref{fig:triggerCounts}, but also demonstrate how different chairs provide challenges to the different interaction methods. Interestingly, for 3D Sketching and Physical tablet the figure has a similar profile across models as opposed to VR Tablet and Virtual Whiteboard.

Finally, at the conclusion of the experiment the user evaluated the different methods of interaction shown in  Figure~\ref{fig:user_experience} is the cumulative score. 3D sketch appears to provide the best user experience. However, the trend is inconsistent with the number of triggers of figs~\ref{fig:triggerCounts} and ~\ref{fig:radar_chart_tries}.

\subsection{Discussion}\label{sec:discussion}

The difficulty in finding some 3D models with 2D techniques in the database becomes evident from Figure~\ref{fig:radar_chart_rs}.
Despite all the methods being intuitive, the 3D sketch is more accurate and satisfying to the user experience.

In the case of 3D Sketch the user can depict the target chair using more naturally the depth information,
while in all the other methods user can draw only a 2D projection of a three dimensional data on a texture.
Intuitively the search results will be improved in terms of precision since more sketch features can be extracted and evaluated by the system.
The real tablet method introduces physical feedback for the user as the draws are generated by the finger when touching the surface, but essentially the output of the interaction is the same for all the 2D techniques.
As each user performs 32 searches in total, they develop a plan to optimize the search, exploring
how the input impacts on the neural network and exploiting eventual winning strategies for each method.

\begin{figure}[t]
     \includegraphics[width=\columnwidth]{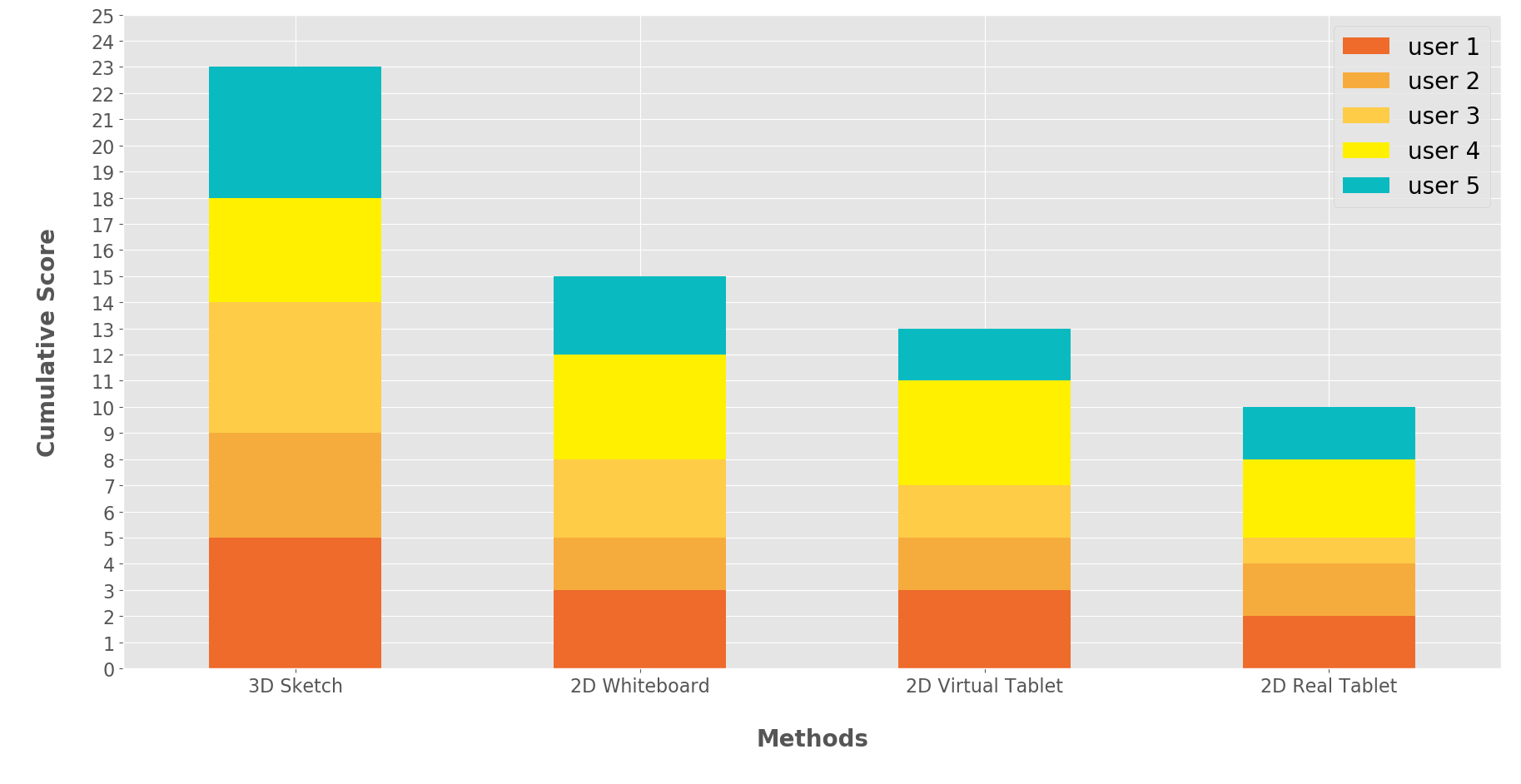}
     \caption{ Each bar is the cumulative score given by each user for a specific method. }
     \label{fig:user_experience}
\end{figure}

\begin{figure}[t]
     \includegraphics[width=\columnwidth]{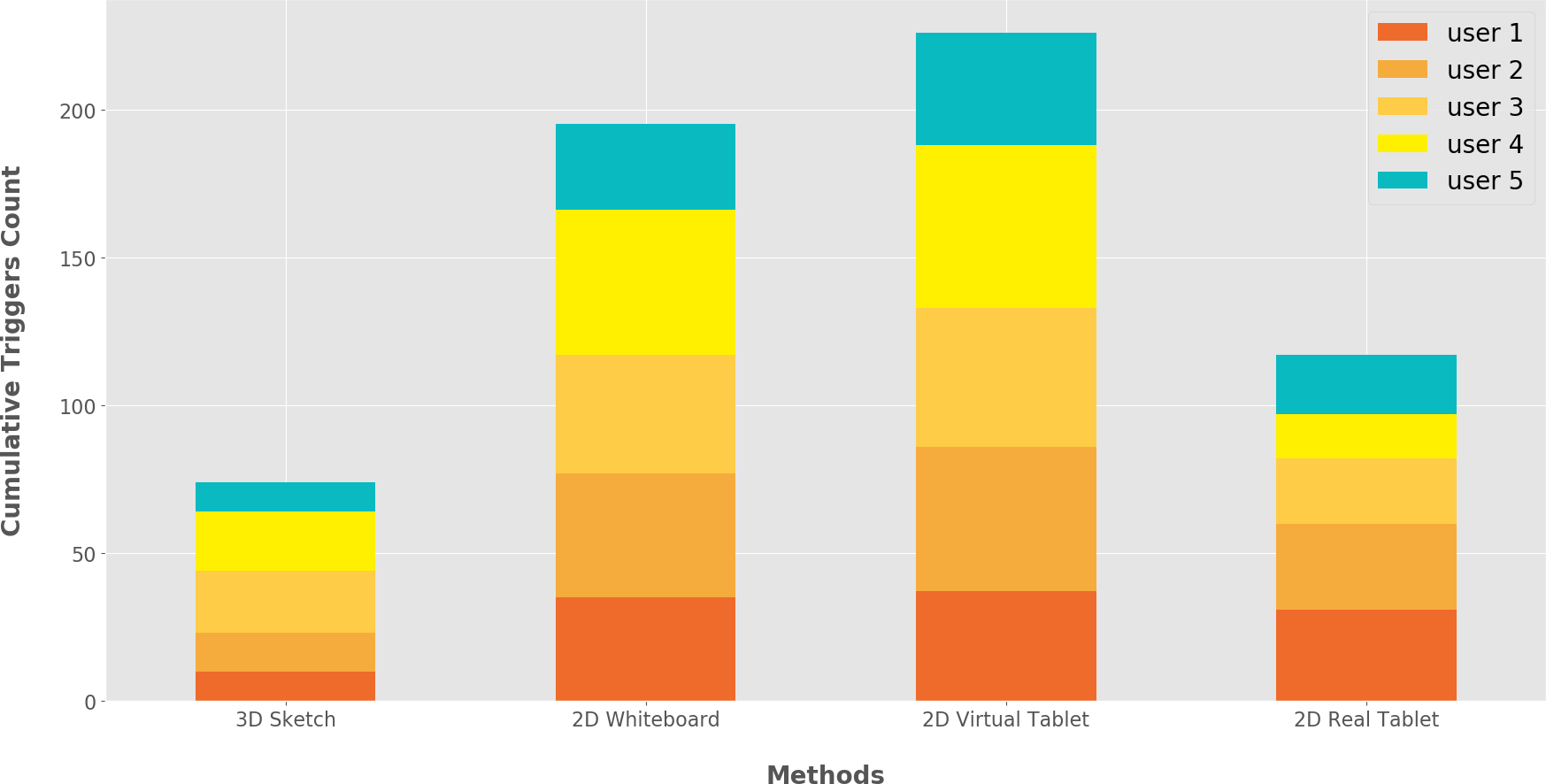}
     \caption{Each bar is the cumulative number of search trigger events given by each user for a specific method. }
     \label{fig:triggerCounts}
\end{figure}

As 3D sketch interaction generates 12 input images from different points of view, the 2D sketch can contribute only with one. With 2D sketch, each user quickly developed the idea of selecting the most significant view angle and tries to depict that projection. Despite this, as the success rate for 2D methods is lower than 3D Sketching. Some users tried to draw the different points of view in the same texture
, in some cases achieving successfully the target chair.
Also, while the 3D sketch does not require an accurate depiction, we noticed that for each 2D methods
the user tended to detail the drawing to increase the probability of finding the target.

\begin{figure*}
    \includegraphics[width=7in]{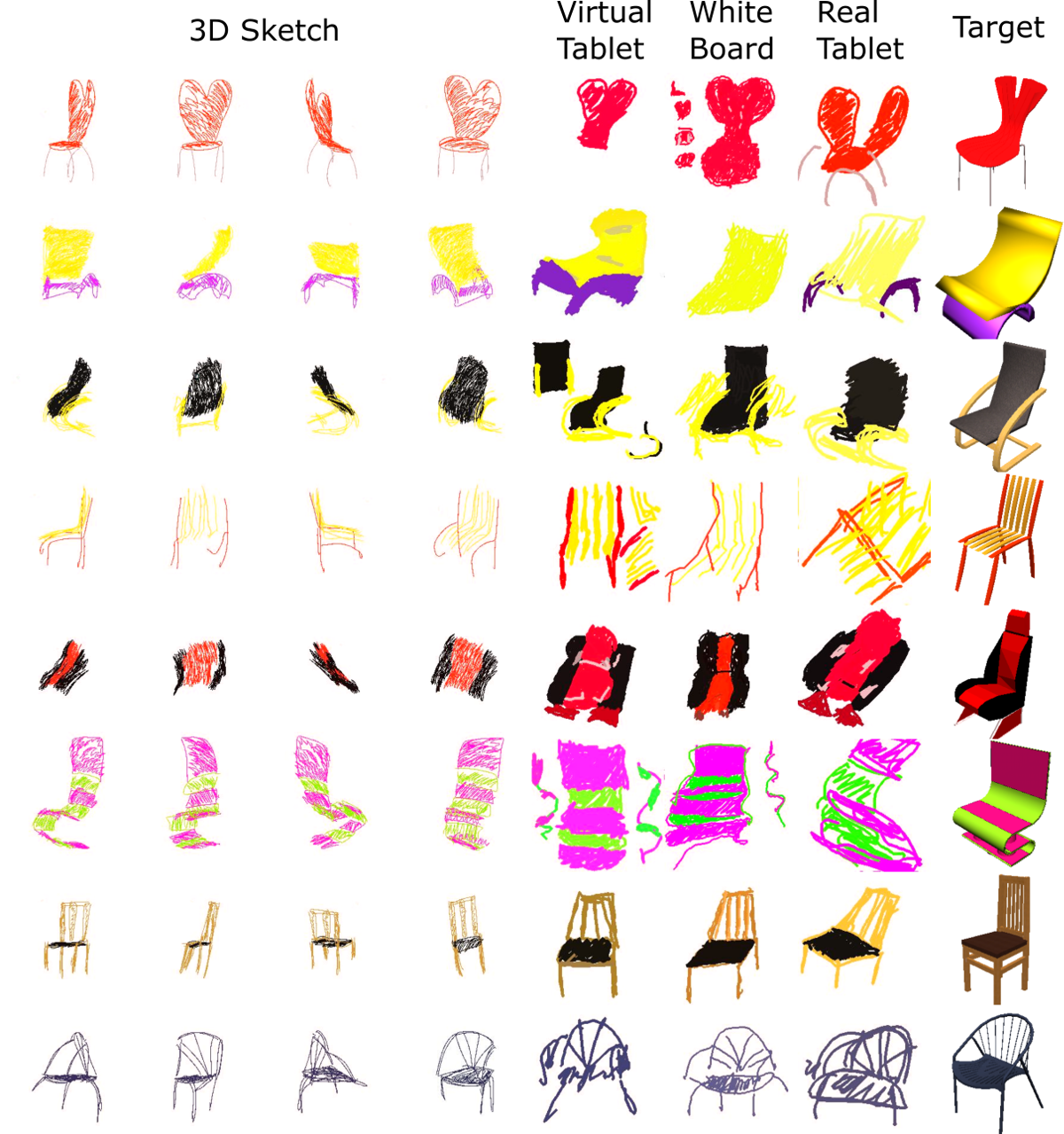}
    \caption{ This figure shows in the first four columns some representative images from the 3D sketch. The fifth column is the sketch from the virtual tablet method. The sixth column is the outcome of the whiteboard method and the seventh column, the real tablet sketch. The last column is the image of the target chair. }
    \label{fig:collagesketch}
\end{figure*}

As a consequence, a typical behavior noticed on the 2D canvas is that the user preferred to fill accurately areas between the edges. This disposition is not necessary for 3D where a few quick strokes can be interpreted by the system as a filled surface.
The motivation for this differences is again the lack of information of 2D sketch compared with the 3D counterpart, so the need to detail the 2D drawing as much as possible emerges naturally.

This lack of efficiency for the 2D sketches harm the user experience. Essentially the 2D drawing is conditioned by a continuous search of the feature that can trigger the right system's response. Therefore, this attitude denatures the process of intuitive and iconographic representation of a model. 

In terms of user interaction, for the aforementioned reasons 2D sketch modalities required from the user more query submission to the system (as shown by \autoref{fig:triggerCounts}), more attention, a firmer hand, and in particular with virtual tablet two hands working at the same time. This could have caused discomfort after a few minutes of sketching and eventually a loss of accuracy.

Between 2D methods, the whiteboard shows better results than the virtual tablet for two main reasons: firstly because of the fixed texture to draw on, and secondly because of the larger canvas that can include more details. Despite the distance between the surface and the drawing hand is the largest between the 2D methods, this result shows that this aspect does not have a negative impact on the task.

The real tablet is interesting because of the physical interaction that is completely absent in all the other methods. While at the beginning of the session this is a pleasant novelty, contrary to what was imagined the sketch done directly with the finger was less precise if compared to the sketch generated by a remote controller. Moreover, mixing virtual reality with a tracked canvas was not sufficient to guarantee decent user experience. The absence of the positional information of the finger lets the user, still immersed in the scene, become disoriented during the drawing process as it is unclear where is the finger respect to the canvas position.
Despite that we managed this issue using a Leap Motion to track continuously the finger displaying an avatar, sometimes noise was introduced and the user experience deteriorated rapidly.
In addition, as it inherited the lack of performance of the 2D techniques, this method does not improve the results of the other 2D methods. This could indicate that even synthetic haptic feedback could aid in the drawing for the other methods even in 3D Sketching.

Both Figure~\ref{fig:triggerCounts} and Figure~\ref{fig:radar_chart_tries} combined with the success rate of the methods motivate the low value of triggering events for 3D sketch and real tablet techniques. 3D sketch had a high rate of success and a low number of queries that indicates the high efficiency of the method. On the contrary, even if the real tablet had few triggering events it does not mean that is efficient. Its low success rate and problematic interaction lead the user to give up quickly, avoiding an extensive search as in the other 2D methods. 

All the aforementioned considerations impact on the user experience as shown in Figure~\ref{fig:user_experience}.

\subsection{Limitations}
\label{sec:limitations}

Despite 3D sketch showing positive feedback from the user study compared to the other methods, we investigate some aspects and limitations of this system to improve the search accuracy or experience. Below we outline the most important ones: 

\noindent{\bf Query Descriptors}
In our study, we used VGG-M deep descriptor, but recently a large number of CNN architectures has been designed guided by the work of Giunchi \etal. We implemented our system as distinct modules that can be replaced easily, threrefore being relatively simple to extend the study to different neural networks and machine learning models in order to make a comprehensive survey. These solutions could also be fine-tuned to VR or learnt through active learning to become bespoke to the users style. We however focus on the user expeirence with the different modalities in this study.

\noindent{\bf Expanded comparisons}
Although our study aimed to compare 4 comprehensive methods for retrieval using sketches in the style of both 3D or 2D, the study could extend the study to other methods of interaction. 
An additional method could be a tracked pen coupled with the tablet to increase accuracy. An interesting follow-up would be comparing the user experience and accuracy obtained in a virtual environment with a real counterpart. In such case the interaction with the physical tablet would likely create a more pleasing user experience.

\noindent{\bf Sketch in Augmented Reality} Understanding how users sketch within a virtual reality allows us to perform controlled user studies, but many application fields require real-world interaction. By trivial adaptations to this method, we could see applications within Film \& TV for set design or architecture for model creation. Such applications require their nuances to be consider and would need to take advantage of the aforementioned extensions.\\

In addition, we aim to compare the sketch mechanism with more advanced user interfaces paired with state-of-the-art mechanisms for searching objects in immersive environments such as text input and faceted search. Moreover, we want to explore possible ways to integrate sketch with additional information coming from a more complex interface. Furthermore, several benefits can be achieved providing functionalities such as brush size, erase and transform operations that could mimic the basic functionalities of 2D photo editing software.

\section{Conclusion}
\label{sec:conclusion}
In the task of scene modeling we have investigated different strategies of interaction within Virtual Reality.
Framed within the context of database navigation and retrieval for scene modeling, sketch interaction has been shown to be a satisfying method when compared with text queries or even the simple database scrolling allowing expressive visual description of the query and accurate results to be retrieved.
In addition, 3D and 2D sketch are extensively studied in different contexts related to database navigation and we have  filled a needed missing comparative study performed within VR.
Our study fills this gap by comparing different sketch-based mechanisms including 3D, 2D sketch in virtual reality with a tablet or a whiteboard and a method that considers the use of a physical tablet while the participant is immersed in the virtual environment.
In our experiment, we collected the time to perform the task, the rate of success, and the number of queries.

We analyzed the methods and discovered that amongst the 2D methods, the provision a physical tablet did not improve the user experience. It is intuitive to conclude 3D sketching as a more suitable user experience within VR, therefore overcoming the familiarity of 2D based retrieval. We therefore pose 3D mid-air sketching as a suitable direction within VR retrieval and related tasks.



\bibliographystyle{ACM-Reference-Format}
\bibliography{sample-base}


\end{document}